\documentclass[useAMS,usenatbib,fleqn]{mn2e}

\pdfoutput=1

\usepackage{txfonts}
\usepackage{ulem}
\usepackage[pdftex]{graphicx}
\usepackage{times}             

\newcommand{\hmsun}{h^{-1}{\rm M}_\odot}
\newcommand{\hmpc}{h^{-1}{\rm Mpc}}

\newcommand{\kms}{{\rm ~km/s}}

\title[Structure and dynamics in void shells]{Clues on void evolution
III: Structure and dynamics in void shells}

\author[Andr\'es N. Ruiz et al.]{\parbox[t]{\textwidth}{\vspace{-1cm}
                                 Andr\'es N. Ruiz$^{1,2}$\thanks{e-mail: andresnicolas@oac.uncor.edu},
				 Dante J. Paz$^{1,2}$,
				 Marcelo Lares$^{1,2}$,
				 Heliana E. Luparello$^{1}$,
				 Laura Ceccarelli$^{1,2}$,
				 Diego Garc\'ia Lambas$^{1,2}$
			         }\vspace{0.2cm}\\
$^{1}$Instituto de Astronom\'{\i}a Te\'orica y Experimental, CONICET--UNC, Laprida 854, X5000BGR, C\'ordoba, Argentina.\\
$^{2}$Observatorio Astron\'omico de C\'ordoba, Universidad Nacional de C\'ordoba, Laprida 854, X5000BGR, C\'ordoba, Argentina.\\
}

\begin{document}

\date{Accepted XXX. Received XXX; in original form XXX}

\pagerange{\pageref{firstpage}--\pageref{lastpage}} \pubyear{XXXX}

\maketitle

\label{firstpage}


\begin{abstract}

Inspired on the well known dynamical dichotomy predicted in voids, where some
underdense regions expand whereas others collapse due to overdense surrounding
regions, we explored the interplay between the void inner dynamics and its
large scale environment.  The environment is classified depending on its
density as in previous works. We analyse the dynamical properties of
void-centered spherical shells at different void-centric distances depending on
this classification.  The above dynamical properties are given by the angular
distribution of the radial velocity field, its smoothness, the field dependence
on the tracer density and shape, and the field departures from linear theory.
We found that the velocity field in expanding voids follows more closely the
linear prediction, with a more smooth velocity field.  However when using
velocity tracers with large densities such deviations increase.  Voids with
sizes around $18\hmpc$ are in a transition regime between regions with
expansion overpredicted and underpredicted from linear theory.  We also found
that velocity smoothness increases as the void radius, indicating the laminar
flow dominates the expansion of larger voids (more than $18\hmpc$).  The
correlations observed suggest that nonlinear dynamics of the inner regions of
voids could be dependent on the evolution of the surrounding structures.  These
also indicate possible scale couplings between the void inner expansion and the
large scale regions where voids are embedded.  These results shed some light to
the origin of nonlinearities in voids, going beyond the fact that voids just
quickly becomes nonlinear as they become emptier.  
\end{abstract}

\begin{keywords}
large--scale structure of the universe - methods: numerical -
methods: statistical
\end{keywords}

\section{Introduction}
\label{sec:introduction}

Voids are prominent features of the large-scale structure of the Universe since
they are surrounded by elongated filaments, sheetlike walls and dense compact
clusters, setting the pattern of the cosmic web \citep{colless_2df_2001,
colless_2df_2003,tegmark_sdss_2004,huchra_2mass_2005}.
Also, voids are likely to determine large-scale velocity fields by their radial
flows of mass and galaxies induced by the local mass underdensity
\citep{bertschinger_self-similar_1985,melott_generation_1990,mathis_voids_2002,
colberg_intercluster_2005, shandarin_voids_2006,platen_alignment_2008,
aragon-calvo_hierarchical_2013,patiri_quantifying_2012,paz_clues_2013,2014_viper_Micheletti}.
The spatial and dynamical statistics of cosmic voids provide critical tests of
structure formation and cosmological models 
\citep[e.g.][]{peebles_void_2001,benson_galaxy_2003, park_challenge_2012,
kolokotronis_supercluster_2002, colberg_voids_2005, lavaux_voids_2010,
biswas_voids_2010,  bos_darkness_2012, park_challenge_2012, bos_less_2012,
hernandez-monteagudo_signature_2013, clampitt_voids_2013}.

The voids in the galaxy distribution have been extensively studied and
characterized by several authors in a variety of wavelengths
\citep{pellegrini_voids_1989,slezak_objective_1993,el-ad_voids_1997,
el-ad_catalogue_1997,aikio_simple_1998,el-ad_case_2000,muller_voids_2000,arbabi_void_2002,
plionis_size_2002,hoyle_voids_zwicky_2002,hoyle_voids_2df_2004,hoyle_voids_sdss_2005,
colberg_voids_2005,ceccarelli_voids_2006,patiri_statistics_2006,shandarin_voids_2006,
platen_voids_2007,brunino_voids_2007,hahn_voids_2007,neyrinck_zobov_2008,
foster_voids_2009,lavaux_voids_2010,tavasoli_challenge_2013,elyiv_voids_2014} and show similar
properties, even though a variety of identification methods and data samples
are used \citep{colberg_aspen-amsterdam_2008}.
Also, the void phenomenon has been a target of many theoretical analysis in
numerical simulations \citep{hoffman_origin_1982,hausman_evolution_1983,
fillmore_self-similar_1984,icke_voids_1984,bertschinger_self-similar_1985,
kauffmann_voids_1991,padilla_spatial_2005,aragon-calvo_unfolding_2010,
aragon-calvo_hierarchical_2013}. 

As voids expand, matter is squeezed in between them and the walls, filaments
and clusters, generating the evolution void boundaries \citep{martel_simu_1990,
regos_void_1991,dubinski_void_1993,van_de_weygaert_voids_1993,bond_filaments_1996,
goldberg_lfvoid_2005,padilla_spatial_2005,cauntun_evolution_2014}. 

Moreover, at early times the large-scale velocity field outflowing from
underdense regions is associated with the linear collapse of overdense regions
and shaped by the tidal influence of the surrounding mass distribution
\citep[Tidal Torque Theory][]{Doro_1970,Peebles_1969,White_1984,Porciani_2002A,
Porciani_2002B}. This process imprints correlations between the inner dynamics 
of haloes and their host filaments and void distribution as well
\citep{Patiri_2006,Paz_2008,paz_alignments_2011,schneider_shape_2012,smargon_align_2012,
vandaalen_align_2012,zhang_align_2013,forero-romero_align_2012}.
However, this interplay is not only originated by the linear growth of initial
fluctuations, but also affected by the vorticity of the cosmic velocity
field which arises in the non linear phase \citep{pichon_vorticity_1999,
laige_swirling_2013, wang_kinematic_2014}.
The antisymmetric component of the velocity deformation tensor also plays a
major role in shaping the cosmic web, producing preferred directions for dark
matter haloes and galaxies \citep{libeskind_velocity_2014}.

In previous works we examined the distribution of galaxies around voids in the
Sloan Digital Sky Survey (SDSS) and performed a statistical study of the void
phenomenon focussing on void environments \citep[][hereafter Paper
I]{ceccarelli_clues_2013} and void dynamics \citep[][hereafter Paper
II]{paz_clues_2013}.
To that end, we have examined the distribution of galaxies around voids in the
SDSS by computing their redshift space density profile (Paper I), and
recovering their underlying real space density and velocity profiles (Paper
II). The real space profiles were inferred through modeling void-galaxy
redshift space distortions on the two-point correlation function (Paper II).
\citet{sheth_hierarchy_2004} presented theoretical foundations that established
a dynamical dichotomy of voids. According to the authors, there are two
distinct void behaviours depending on their environment: the so called
``void-in-void'' and ``void-in-cloud'' process.  
The void-in-void regions are embedded in larger-scale underdensities and show
expanding velocity profiles.
On the other hand, void-in-cloud regions are surrounded by larger overdense
environments which undergo in a former collapse, shrinking at later times the
embedded void region. 
This last case seems to affect more likely small rather than large voids.  
The main goal of our previous studies was to confirm by the first time on
observations this dynamical classification.

We defined in Paper I a separation criterion based on the redshift space
integrated density profile to characterize voids according to their surrounding
environment. Such criteria have shown success in separating expanding and
contracting voids (Paper II) and allow us to define two characteristic void
types: (i) voids with a density profile indicating an underdense region
surrounded by an overdense shell, dubbed S-type voids, which behaves like a
``void-in-cloud'' process; and (ii) voids showing a continuously rising
integrated density profiles, defined as R-type voids, which expand in a similar
way as ``void-in-void'' regions.
In this paper we analyse in detail the structure and dynamics of the velocity
field in the regions surrounding voids. 
The aim of this work is to shed some light on the origin of the observed
deviations of void velocity profiles from linear theory (Paper II), and address
whether such deviations arise in the inner shell structure (i.e. the features
of the density field inside shells) or they are more related with the surrounding
large scale characteristics (i.e. void of interest classified as R or S-type).  
To this end we follow our previously defined classification scheme for void
environments (R and S-types), where we analyse the velocity and density field
inside void centric shells.  

The organization of this paper is as follows.
In Section \ref{sec:data} we describe the numerical $N$-body simulation used
and the construction of the corresponding void catalogue. We describe the methods and
present the results of the analysis of the dynamics of void shells in Section 
\ref{sec:pixels}.
In Subsection \ref{sec:shells} we present the analysis of void shells on the
basis of an equal area pixelization scheme, distinguishing pixels according to
their local density.
The kinematics of regions defined by the different sets of pixels are are
analyzed in Section \ref{sec:shells_dynamics}, where we use these sets to
characterize the network structure and the stratum of low densitiy regions.
A comparison of the structures defined by the different sets is made on the
basis of the properties of minimal spanning trees on Sections
\ref{sec:structures} and \ref{sec:structures_dynamics}.
Finally, in Section \ref{sec:conclusions} we present the conclusions of our
results.


\section{Void catalogue}
\label{sec:data}

\begin{figure*}
\centering
\includegraphics[width=0.9\textwidth]{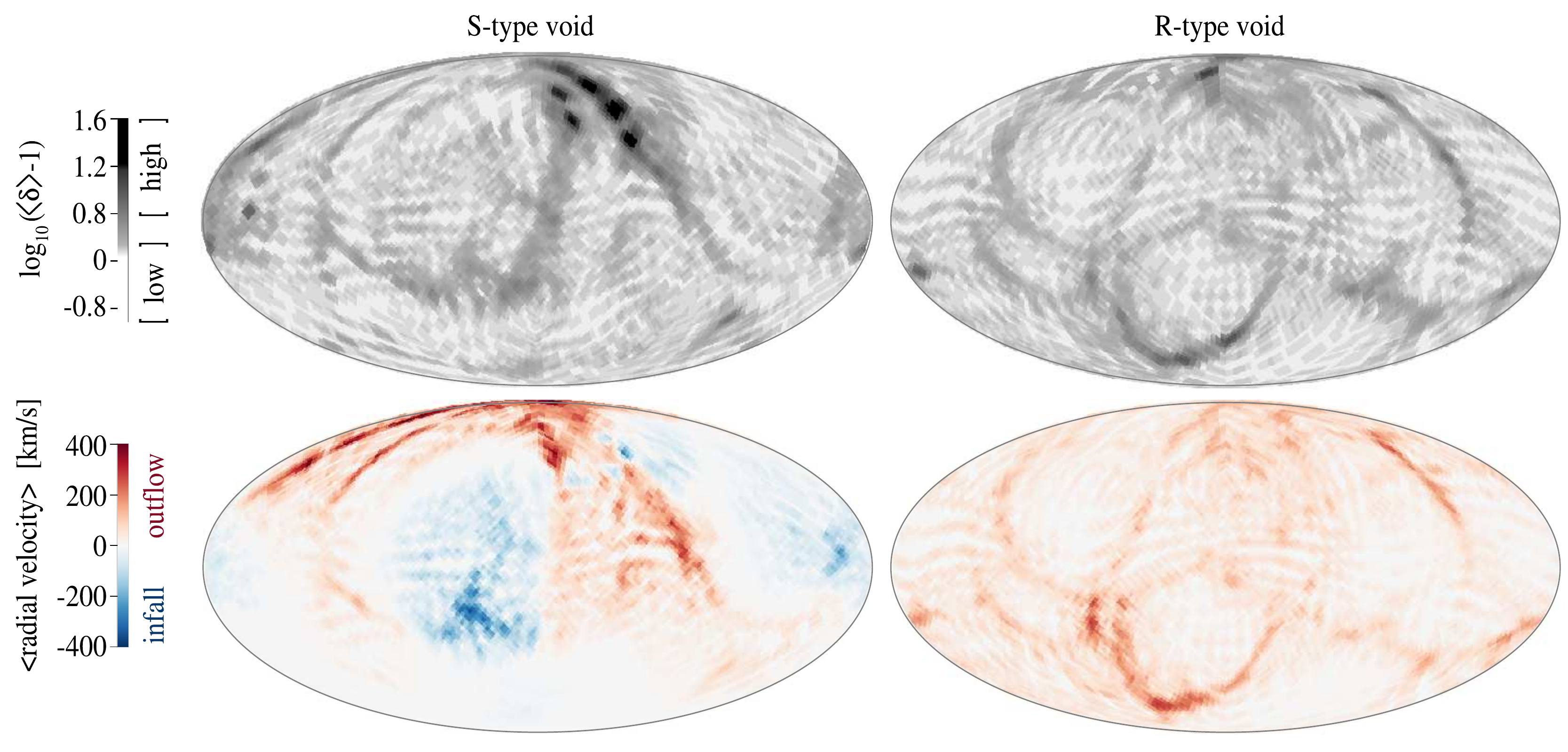}
\caption{
Mollweide projection of mass density contrast (top panels) and radial velocity
(bottom panels) pixel maps of two voids with the utmost S (``void--in--cloud'',
left panels) or R-type (``void--in--void'', right panels) behaviour. All
simulation particles in the range $0.9<r/R_{\rm void}<1.1$ are used to compute
averaged values within each pixel. The two voids shown in this Figure have been
chosen so that they show the greatest separation in the void classification
scheme, based on the integrated density profiles (detailed in Paper I).  This
two cases exemplify how the structure of the velocity field in the inner parts
of a void at $1$ void radius exhibits very distinct behaviour depending on their
environment (defined around $3$ void radius, i.e. R- or S-type void). Such
differences are not expected in linear theory, given that both regions by
definition enclose the same amount of integrated overdensity $\Delta(R_{\rm
void})=-0.9$.
}
\label{fig:mapa}
\end{figure*}

\subsection{The $N$-body simulation} 
\label{sec:simu}

In this work we use a dark matter simulation of $512^3$ particles
evolved from an initial redshift of $z \sim 50$ to the present time in a
comoving box of side length $L=500\hmpc$.  The cosmological parameters
correspond to a flat $\Lambda$CDM model consistent with the WMAP7 estimations
\citep{jarosik_wmap7_2011}, with a matter density parameter $\Omega_{\rm m} =
1.0 - \Omega_\Lambda = 0.258$, a dimensionless Hubble constant $h=0.719$ and a
normalization parameter $\sigma_8=0.796$. The resulting mass particle is
$m_{\rm p} = 6.67 \times 10^{10} \hmsun$.
The initial conditions were generated using \textsc{GRAFIC2}
\citep{bertschinger_grafic2_2001} and the simulation was evolved using the
public version of \textsc{Gadget-2} \citep{springel_gadget2_2005}.  The
identification of haloes was performed using a Friends-of-Friends algorithm
\citep[][hereafter FoF]{huchra_groups_1982, davis_fof_1985} with a percolation
length given by $l=0.2~\bar{n}^{-1/3}$, being $\bar{n}$ the mean number density
of dark matter particles. The final catalogue contains $505126$ haloes with at
least $10$ particles.

\subsection{Void identification}
\label{sec:voidcat}

We perform the identification of voids using a modified version of the
algorithm presented by \citet{padilla_spatial_2005} and
\citet{ceccarelli_voids_2006}.  The identification is done according to the
following steps:
\begin{enumerate}
\item
Using the halo catalogue as structure tracer, we perform a Voronoi tessellation
using the public library \textsc{voro++}\footnote{http://math.lbl.gov/voro++/}
\citep{rycroft_voro++_2009}. This allow us to estimate the density field as the
inverse of the Voronoi volume for each cell.

\item 
The candidates to underdense regions are selected at the centroid position for
each cell with estimated overdensity satisfying $\delta < -0.8$.

\item
Centred on each candidate position, we compute iteratively the integrated
density contrast ($\Delta$) inside spheres of increasing radius, denoted by
$r$. When the integrated overdensity satisfy $\Delta(r) > -0.9$, the iteration
ceases and the current sphere radius $r$ is defined as the radius of the void
candidate.  If the threshold on $\Delta(r)$ is never achieved the candidate is
not considered anymore as a possible void.

\item

Once underdense regions are identified, the step (iii) is repeated starting
this time from a randomly displaced center instead of the candidate centroid.
Such displacement takes the form of a random jump proportional to the void
radius.  These random jumps are performed several times, obtaining a random
walk around each candidate.  Each jump is only accepted if the new obtained
radius is larger than the last accepted value. If the current step is accepted,
the candidate center is updated to the new position.  The purpose of such
random walk is to recenter the identified voids in the largest scale local
minimum of the density field. This procedure derive a well defined center which
allows the computing of several stacking statistics like the correlation function
or radial averages profiles \citep{ceccarelli_large-scale_2008}.

\item
The last step consists on rejecting all the overlapping spheres, starting for
the largest candidate. 
\end{enumerate}

As any void definition, the one presented in the above paragraph have some
features which can be related by analogy to other identification algorithms.
However in order to compare results among the literature corpus some caveats
should be taken into account depending on the details of the void definition.
For instance, several algorithms involve the construction of a hierarchy of
voids, defined as regions enclosed by matter ridges \citep[see for
instance][and references therein]{colberg_voids_2005, platen_voids_2007,
neyrinck_zobov_2008, elyiv_voids_2014}. In those cases, the value of the
density field inside void regions could vary in a wide range. In our case we
select voids by ensuring an enclosed density below a given threshold.  For this
reason, it is not straightforward to relate the underdense regions identified
on both methodologies.

Following the procedure presented in Paper I, we have defined two types of
voids depending on their integrated density contrast profiles.
The mean of these profiles have been computed for different void size ($R_{\rm
void}$) intervals.
As shown in Paper I, such average curves have a well defined maximum at a
distance $r_{\rm max}$ from the void centre, except for the largest voids which
exhibit an asymptotically increasing profile. This radius typically reach void
centric distances around $r_{\rm max}\approx3 R_{\rm void}$.  Therefore the
density at such scales can be thought as a measure of the large scale
environment where the void is embedded.

We classify voids into two subsamples according to positive or negative values
of the integrated density contrast at $r_{\rm max}$. Due to the linear theory,
such definition implies that the region surrounding the void has negative or
positive velocity divergence at those scales. This imply that this region is
either in contraction or expansion, respectively.
Voids surrounded by an overdense shell are dubbed S-type voids, and satisfy the
criterion $\Delta(r_{\rm max}) > 0$.
On the other hand R-type voids are defined as those that satisfy the condition
$\Delta (r_{\rm max}) < 0$, which corresponds to voids with continuously rising
integrated density profiles.
More details about this void classification scheme can be found in Paper I and
II.
The simulation box is found to contain $1622$ voids with sizes ranging from
$\sim9.5\hmpc$ to $\sim29\hmpc$, where $810$ are S-type and $812$ are R-type. 


\section{Analysis of the structure in spherical void-centric shells}
\label{sec:pixels}

In this Section we aim to investigate whether there is a relation between the
spatial distribution, shapes and dynamics of structures surrounding voids,
depending on its classification (i.e. R- or S-type void).
As we summarized in the previous section and we have seen on Paper I and Paper
II, this classification scheme successfully reflects the dichotomy on the
dynamics of underdense regions. 
The linear theory predicts the future collapse and expansion of S- and R-type
voids, respectively.
The theory also predicts that the behaviour in the inner parts of an underdense
region is insensitive to whether it is surrounded by any kind of environment
(i.e., S- or R-type): the velocity profile only depends on the amount of
enclosed mass.
However in Paper II, we have seen significant deviations from the linear
predictions in the inner parts of void velocities profiles, both in R- and
S-type voids (see Figure 3, in Paper II).
Similar qualitative results can be found in the work of
\citet{hamaus_profile_2014}, who found deviations from the predicted radial
velocity profiles of stacked voids in the inner regions (around $0.5$ to $1$
void radii), being these differences more relevant for small voids.
Nonetheless, only a qualitative comparison between our results and those in
\citet{hamaus_profile_2014} is possible, due to differences on the void
definition and identification algorithms.
Voids used in their work are identified accordingly to a hierarchy
\citep{neyrinck_zobov_2008}, where small underdense regions can be found inside
of larger voids with a higher cumulative mass overdensity.
In consequence, there are differences between definitions of void radius and
center. 
At the same time, their results are not classified into S- or R-types and
therefore their stacked profiles reflect the two dynamical
behaviours (specially for small sized voids).

At first glance, it could be conjectured that inner parts of underdense regions
quickly reaches the lower bounds of the validity domain of the linear theory
(i.e. nonlinear dynamics becomes increasingly significant as $\delta$
approaches to $-1$).
However, the purpose of this Section is to get a deeper insight on the issue,
by attempting to derive the specific source of such discrepancies from linear
theory.
To this end, we analyze the properties of different tracers used to map the
void velocity field.
The properties taken into account involve local density, shapes, and spatial
distribution of matter around voids.

We consider spherical shells of
various sizes centered on each void in the sample.
In all cases the radial width of a given shell is a $20$ per cent of the void
radius.
For each void, the smaller shell is centered at $r = R_{\rm void}$ and the
larger one at a distance of $r = 3R_{\rm void}$.
The shells are spherically tessellated with equal area pixels computed using
the software \textsc{HEALPix} \citep{gorski_healpix_2005}.
We need a pixelization scheme where pixels are not too large, in order to
resolve the shapes of structures conveniently. 
However, the shot noise increases when the size of the pixels drops down, since
most of the pixels contain no particles. 
We use a pixelization scheme with 5 levels of refinement ($N_{\rm side}=32$)
giving a total of $N_{\rm pix}=12~N_{\rm side}^2=12288$ pixels for each map.  
The results shown in this work do not depend on the resolution of the
pixelization scheme.

\subsection{Velocity and density angular distribution}
\label{sec:shells}

\begin{figure*}
\centering	
\includegraphics[width=0.9\textwidth]{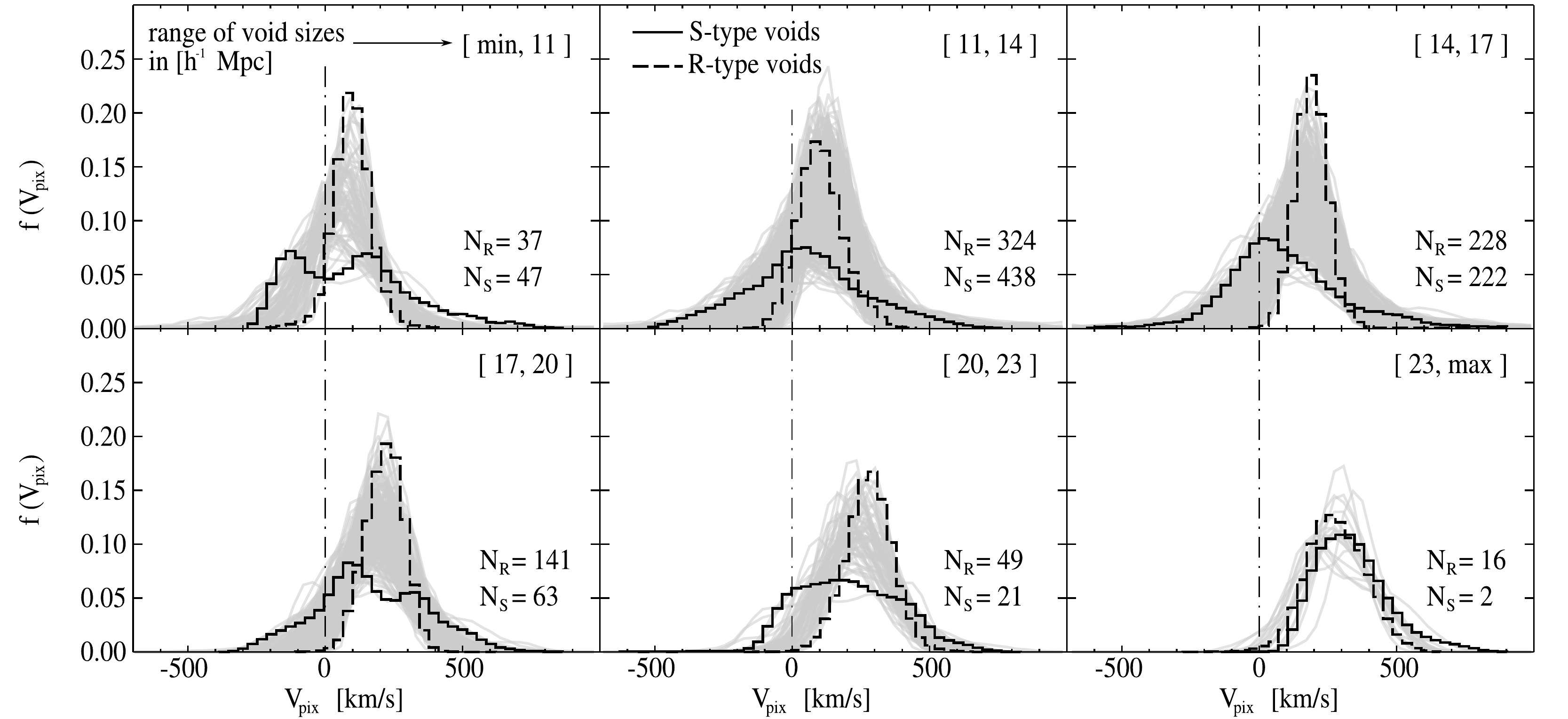}
\caption{
   Normalized histograms of pixel radial velocities for the S-type
   (black solid lines) and R-type voids (black dashed lines) with the largest
   differences in their radial profiles (i.e. the S-type void with the largest
   density around $3R_{\rm void}$ and the R-type void with the lower density at
   the equivalent scale), selected in six void size intervals of width $3\hmpc$,
   except the first and the last ones. Grey curves correspond to all the voids in
   that size range. The dot-dashed line marks the position for $V_{\rm pix}=0$,
   the transition from outflow to infall behaviour. The void size increases from
   top to bottom and from left to right, as indicated in the top-right label of
   each panel. The number of voids with R or S type environments (${\rm N}_{\rm R}$ and ${\rm N}_{\rm S}$) are indicated 
   at the left bottom legends.
}
\label{fig:vpix}
\end{figure*}

We applied the described shell tessellation scheme to all voids in the
simulation box (Sec. \ref{sec:voidcat}) and computed the density and mean
radial velocity of particles on each pixel of a given shell ($\rho_{\rm pix}$ y
$V_{\rm pix}$).
Since this pixelization is performed on an equal area basis, the density is
simply proportional to the number count of particles in a pixel, at a distance
within the limits defined for each shell.
In order to study the dynamics of structures in an expanding/collapsing void
scenario (as discussed in Paper II) we computed the averaged void--centric
radial velocities of the particles per pixel and per void.  
Examples of the pixelized maps of density contrast and radial velocity
corresponding to a shell located at $r=R_{\rm void}$, are shown in Mollweide
projection in Figure \ref{fig:mapa}.
Left panels show the maps for a S-type void, and right panels correspond to a
case of a R-type void.
We selected the S-type (R-type) void with the maximum (minimum) value of
integrated density profile at $r_{\rm max}$ of the void sample (as defined in
section \ref{sec:voidcat}), so that they exhibit the utmost R- and S-type behaviours.
This selection of the utmost R- and S-type void behaviours
has been carried out in $6$ void subsamples in voids size intervals
of $3 \hmpc$ (see Figure \ref{fig:vpix}).
For simplicity, in Figure \ref{fig:mapa} we only show the results for one of
these void subsamples.
The upper panels show the logarithmic density contrast of the mass
$\log_{10}(\delta_{\rm pix}+1) = \log_{10}(\rho_{\rm pix}/\bar{\rho})$, where
$\bar{\rho}$ is the mean density of the simulation. 
In the bottom are the radial velocity maps, showing the infalling and expanding
regions. 
The grid patterns seen in low density regions are due to the relatively small
size of the pixelization scheme chosen, and arise as a result of the simulation
grid. 
However, the selected map resolution results adequate to analyse structures at
high density regions.
By definition of the void identification method, both void shells at $r=R_{\rm
void}$, enclose the same amount of integrated density contrast $\Delta(R_{\rm
void})=-0.9$. 
Even more, both voids have similar sizes, $12.6$ and $15.1\,\hmpc$ for S- and
R-type void respectively. 

The motions of shells enclosing an overdense or underdense sphere arise
naturally as a consequence of the gravitational pull and is predicted in its
simplest form by the linear theory \citep{peebles_peculiar_1976}.
According to this, in a standard $\Lambda$CDM model, the radial velocity of a
shell at radius $r$ enclosing an integrated mass overdensity $\Delta_{\rm m}$
is given by:
\begin{equation} V_{\rm lin}(r) = -{H \over 3} r\Delta_{\rm m}(r) \Omega_{\rm
m}^{0.6}, \label{eq:vlin} \end{equation}
where $H$ is the Hubble parameter and $\Omega_{m}$ is the cosmological dark
matter density parameter.
At first glance, by Eq. (\ref{eq:vlin}) we roughly expect the same behavior of
the mean radial velocities for the two cases shown at  Figure \ref{fig:mapa} 
(i.e. by definition both enclose almost the same amount of matter density).
However, as it can be seen in the lower panels of Figure \ref{fig:mapa}, the
fluctuations of the velocity field around this mean velocity behaves notably
different. 
In the S-type shell there are regions with net infall (blue pixels) or outflow
velocities (red pixels). 
In contrast, for the R-type spherical shell we only observe positive radial
velocities (i.e., outflow). 
Also, by comparing bottom to top panels it can be seen that the regions with a
defined sign in radial velocity (i.e. outflow or infall) seem to correlate with
projected structures in the density map.
We have found similar features in all of the $6$ void size subsamples mentioned
above.
Even though, we should emphasize that the maps shown in Figure \ref{fig:mapa}
are not the typical behaviour of S- and R-type voids. 
As stated above, they are just representative of voids with the largest
environment differences.
Since our void classification scheme is based only on two categories, there are
plenty of voids with density profiles which fall into a grey area between both
cases. 
However, guided by the insights obtained here, in the following Sections we
will perform a statistical analysis intended to characterize the features of
the angular velocity/density field in void centric spherical shells.

In order to complement the discussion based on Figure \ref{fig:mapa}, we also
analyse the utmost R- and S-type voids (as defined previously) in six void size
ranges.  For this purpose, in Figure \ref{fig:vpix} we plot the histograms of the
pixel averaged radial velocity for these voids, showing S-type in black
solid lines and R-type in black dashed lines.  The results for different sizes
are shown on each panel as indicated in the top-right labels. 
Grey curves display the corresponding histograms for all voids in a given size range.
As it can be seen, in all cases the utmost R-type voids present a narrow velocity distribution
with a well defined peak, and with almost all the pixels with a positive radial
velocity (i.e. outflow behaviour). 
In contrast, the utmost S-type voids show wider distributions and in a variety of
shapes, covering from flat to multi peak distributions, and with a large
component of infalling pixels (i.e. negative radial velocity). 
As can be seen, the velocity histograms for all voids seem to be enclosed by
the velocity distribution of the extreme R- and S-type voids.
This indicates that the selection of utmost behaviours based on its density profiles
is also manifested in the fluctuation of the velocity field in the inner parts of voids 
(the shell is located at $r=R_{\rm void}$).
For the last size interval (largest voids, lower-right panel), S- and R-type
voids show similar distributions. 
This is expected because larger voids tends to have an R-type behaviour
and the fraction of S-type voids decrease (see Paper I for details).

The results presented in this subsection suggest a systematic correlation
between the fluctuations of the velocity field at the inner parts of voids
(i.e. $r=R_{\rm void}$, where the integrated density is around $\Delta=-0.9$)
and the void environment at large scales ($r\sim 3R_{\rm void}$). 
As discussed at the beginning of this Section, these correlations can shed some
light on the origin of nonlinearities in the inner parts of the underdense
regions. 
This kind of systematic deviations in the velocity field may have in
principle some impact on studies using void-galaxy redshift space statistics
\citep{lavaux_voids_2010,paz_clues_2013,Hamaus_cosmology_2014}.  
The fact that the regions of infall or outflow on the utmost S-type voids seem
to be clustered in angular positions (see left-lower panel in Figure
\ref{fig:mapa}), indicates that it may be a correlation between the velocity
deviations around the mean flow and the properties of the large scale
structures.  
Inspired on these features, in the following sections
(\ref{sec:shells_dynamics} and \ref{sec:structures_dynamics}) we will analyse
the dependence of the velocity field on the density and shape of the surrounding
structures.

\begin{figure*}
\centering
\includegraphics[width=0.9\textwidth]{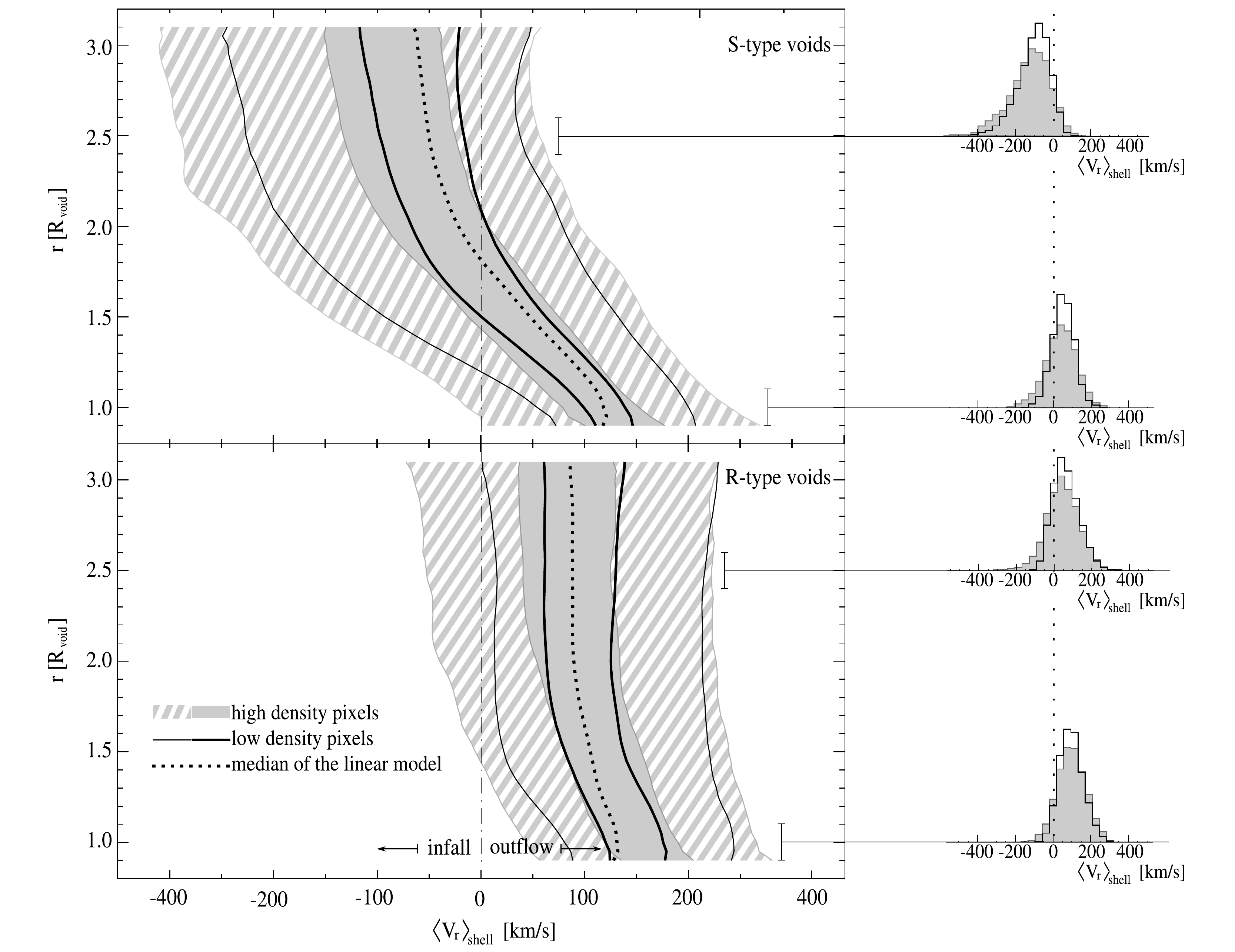}
\caption{
Joint probability density distribution of shell mean radial velocity $\left<
V_{\rm r} \right>_{\rm shell}$, and the shell radius $r$ (in void size units
$R_{\rm void}$), for S-type (top panel) and R-type (bottom panel) voids.  Solid
lines show isopercentiles of the distribution of low density pixels and shaded
regions show the same percentiles corresponding to network structures, formed
by high density pixels. Inner curves/regions represent the $40$ per cent of the
sample and the outer ones the $90$ per cent. The dotted central line
corresponds to the median of the linear model prediction given by Eq.
(\ref{eq:vlin}) and the dot-dashed line marks the position of $\left<V_{\rm r}
\right>_{\rm shell}=0$, i.e. the infall/outflow transition. Plots on the
right-hand side of Figure show the normalized histograms of radial velocities
for two intervals of radial distance: a shell at nearly one void radius and a
shell at roughly $2.5$ void radius. The dotted line that crosses the histograms
indicates the position of $\left< V_{\rm r} \right>_{\rm shell}=0$.
}
\label{fig:vr_r}
\end{figure*}

\subsection{Dynamics of low and high density regions in void shells}
\label{sec:shells_dynamics}

As described in the previous section, for each void we compute angular maps of
radial velocity and density in spherical void centric shells of varying radii.
Shell radii span a range between $1$ and $3$ times the void radius.
We divided the full set of pixels in each shell into three categories according
to its density contrast with respect to the mean density of the simulation
($\delta_{\rm pix}$): low density regions are defined by pixels with
$\delta_{\rm pix} < 1$, intermediate density regions satisfy $1<\delta_{\rm
pix}<5$ and high density regions are selected with $\delta_{\rm pix}>5$.  
Such selection is intended to separate high and low density structures from the
stratum of mass with an average behaviour.
For each shell we obtain the average radial velocity (denoted by
$\left<V_r\right>_{\rm shell}$) by computing the average of pixel radial
velocities in one of the three density subsets defined above.
This allows to roughly estimate whether structures of a given density are
moving away from or falling into the void center (i.e. the shell pixel subset
of a given density have a net expansion or contraction).
Finally, following Eq. (\ref{eq:vlin}), we calculate the linear prediction of
radial velocities, $V_{\rm lin}$, for all voids at the different shells.

In Figure \ref{fig:vr_r} we show the distributions of these average velocities
as a function of the shell radius normalized to the void radius. 
These distributions represent the joint probability density of shells and were
estimated by simply counting high or low density shells in bins of $r$ and
$\left<V_r\right>_{\rm shell}$.
Since by definition we have two velocity averages (high and low density pixel
subsets) at all shell radii for each void, the sampling in shell sizes is
homogeneous.
This allows to directly compare between the results corresponding to each
subset centered rather on R- or S-type voids, and also to perform comparisons
between both void categories.
Solid lines in Figure \ref{fig:vr_r} show isopercentile curves of the
distribution corresponding to the subset of low density pixels, whereas shaded
regions indicate the same percentiles for the subset of high density pixels.
The inner curves/regions represent the $40$ per cent of the sample and the 
outer ones the $90$ per cent.
The dotted central line corresponds to the median of the distribution of the
linear model prediction.
The top and bottom panels show the results for shells centered on R- and S-type
voids, respectively.
On the right-hand side of Figure we display histograms of radial velocities
for two intervals of radial distance: a shell centered at one void radius and a
shell at $2.5$ void radius. Finally, the vertical dotted line indicates the
position of $\left<V_r\right>_{\rm shell}=0$ in the histograms. 

It is worth noticing that in Figure \ref{fig:vr_r} the mean radial velocity
distributions for all cases (R- or S-type voids with high or low density
pixels) qualitatively follow the behaviour of the linear theory prediction.
In all cases this model is enclosed within the central $40$ per cent of the
distributions.
This is consistent with the previous results reported on Paper II, where we
showed that linear theory can be used to model redshift distortions around
voids.
In the same direction, S-type voids exhibit typical negative velocities at
large shell radius (around $1.5$ to $3$ void sizes). 
This is the typical behaviour of ``void-in-cloud'' regions described by
\citet{sheth_hierarchy_2004} and observed on the SDSS (Paper II).  
On the other hand, by looking at the top panel of Figure \ref{fig:vr_r} (S-type
voids), it can be seen that the velocity distribution of low density pixels is
nearly symmetric around the linear theory median prediction. 
In contrast, high density pixels exhibit a shift towards greater infall
velocities with respect to the median linear theory. 
Also the outer contours (which enclose the $90$ per cent of the population)
indicate a broader distribution of velocities at these densities. 
This suggests that the larger deviations from linear theory observed on S-type
voids (Paper II) are mainly reflected in the dynamics of the surrounding high
density peaks.
However, we should emphasize that the deviations described here are around the
median linear theory values. 
These deviations are source of dispersion and errors in stacking statistics as
those presented in Paper II, where the dynamics of a given void sample is
modeled by applying linear theory to the mean or median density profile.
Studies using a mixture of S- and R- type voids \citep[see for
instance][]{hamaus_profile_2014}, have also report systematics deviations from
linear theory, in qualitative agreement with our results.
For the case of R-type voids (bottom panel), it is worth noticing that more
than $90$ per cent (outer contours) of low density pixel averages distributes
only over positive velocities. 
According to the ``void-in-void'' dynamics, this type of voids should only
display expansion velocity curves.  
However, the high density pixels show a significant fraction of structures with
net infall averaged velocities. 
Here again the high density peaks at distances ranging from $\sim1.5$ to
$\sim3$ void radius are the main responsible for the deviations from linear
theory. 

\begin{figure}
\centering
\includegraphics[width=0.45\textwidth]{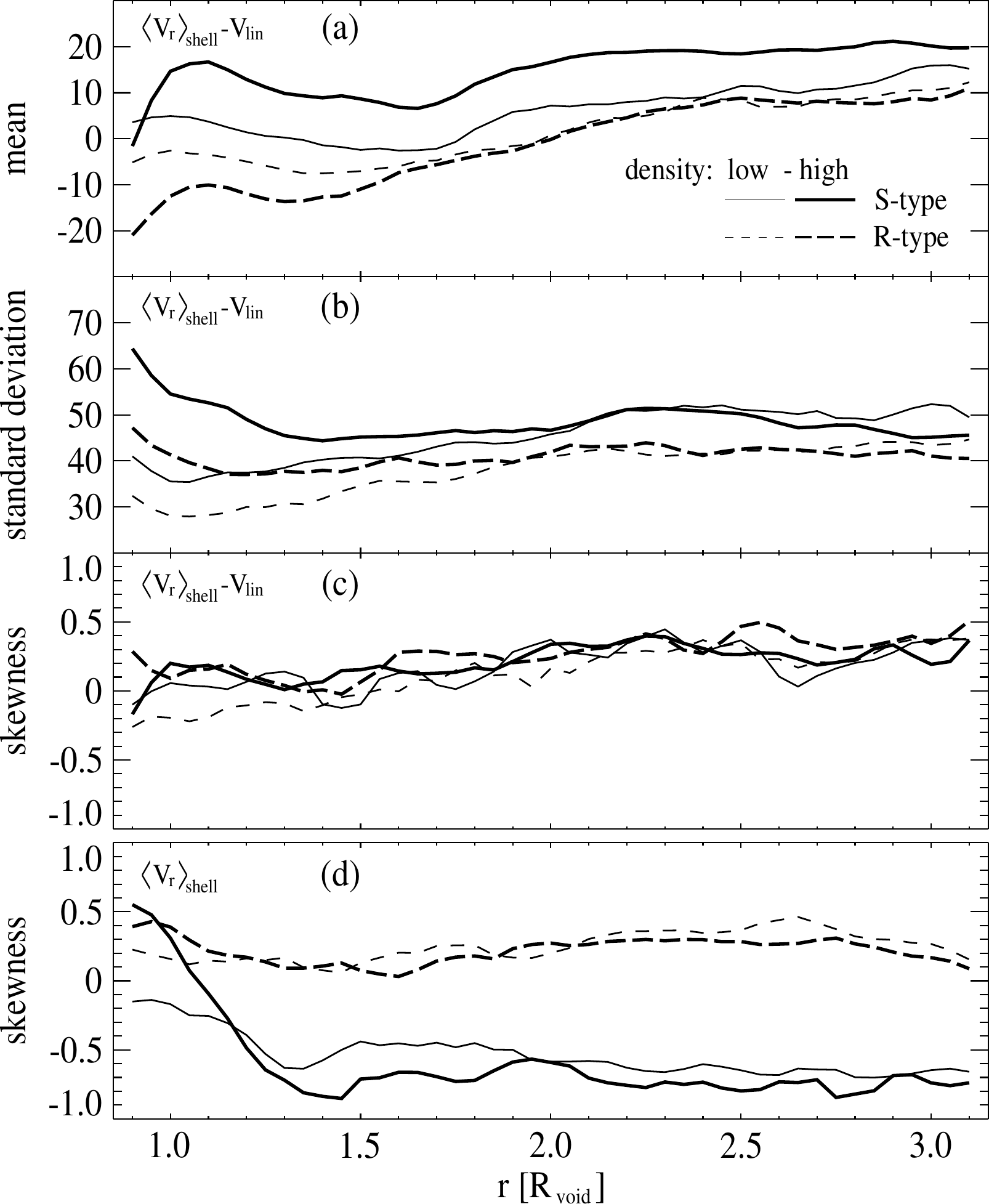}
\caption{
Moments of the distributions of radial velocity relative to the linear
prediction ($\left<V_r\right>_{\rm shell}-V_{\rm lin}$) as a function
of the void--centric distance (three first panels from up to bottom).
S-type voids are indicated with solid lines and R-type voids with dashed lines.
The high density regions are plotted with the thick lines and the low
density ones with thin lines. In the fourth panel (bottom), it is displayed
the third standardized moment (skewness) for the same distributions of radial
velocities of Figure \ref{fig:vr_r}.
}
\label{fig:moments}
\end{figure}

We have analysed deviations of the velocity field from the median of linear
theory predictions depending on the density of the pixel and the type of the
void centre.
In order to see whether such deviations relies in the first order approximation
of linear theory, we study the distribution of radial velocity in void
shells relative to its corresponding linear prediction.
To this end, for each shell and pixel density subset, we compute the difference
between the mean radial velocity and the value expected from linear theory
(i.e.  $\left<V_r\right>_{\rm shell}-V_{\rm lin}$).
Looking for a more quantitative analysis, for all distributions we compute four
moments: the first and second moments (mean and standard deviation) plus third
and fourth standardized moments (skewness and kurtosis).  
The results are shown in Figure \ref{fig:moments}, where S-type voids are
indicated with solid lines whereas R-type voids are drawn with dashed lines.
The thickness indicates low and high density regions for thin and thick lines,
respectively.
The first three panels from up to bottom correspond to the mean, standard
deviation and skewness, whereas the kurtosis is not show because it is
statistically consistent with zero at all radii.
In the panel (d) , we show the skewness of the distributions of
$\left<V_r\right>_{\rm shell}$. 
As can be seen, the results shown in this panel are consistent with
distributions displayed in Figure \ref{fig:vr_r}.
The S-type void velocity distributions show asymmetry toward infall velocities
(negative skewness) for both high and low density regions.
Such asymmetry could arise from the systematic collapse of the surrounding void
region.
When the linear velocity is subtracted (panel (c)), the skewness of the
distributions vanishes or diminishes considerably. 
This indicates that the deviations from gaussianity in the velocity
distributions can be explained mostly by means of linear theory.
On the other hand, R-type voids exhibit marginally positive skewness in its
velocity distribution (panel (d) in Figure \ref{fig:moments}), which is
consistent with the observed suppression of negative velocities in Figure
\ref{fig:vr_r}.
Respecting to the mean of velocity deviations (panel (a) in Figure
\ref{fig:moments}), S-type voids show larger deviations from linear theory than
R-type voids, along the whole scale range.
Among S-type voids, the largest deviations are present when using high density
regions to compute radial velocities.
However at shell radii below $1.5$ void sizes, R-type voids also present
important deviations from linear theory values. 
These results suggest that low density tracers for S-type voids and high
density tracers for R-type voids are more likely to reproduce linear theory
predictions.
This can be important in studies that depend sensitively on the modeling of the
velocity field surrounding voids. 
As an example, it could be mentioned works intended to perform the
Alcock-Paczynski test \citep{2012Lavaux,sutter_first_2012} on voids identified
on sparse galaxy samples, which are expected to be dominated by R-type voids
(Paper I) and high luminosity tracers. 
These studies are sensible to
anisotropies induced by redshift space distortions (Paper II), and therefore
a proper modeling of the velocity field is required. 
In the same direction, studies based on low redshift samples, as the main
galaxy sample of the SDSS, can be benefited by an appropriate treatment of the
velocity tracer population. 
The last rely in the fact that such samples are expected to be dominated by
S-type at small size voids whereas at larger size R-type voids become more
frequent (Paper I).

\begin{figure}
\centering
\includegraphics[width=0.45\textwidth]{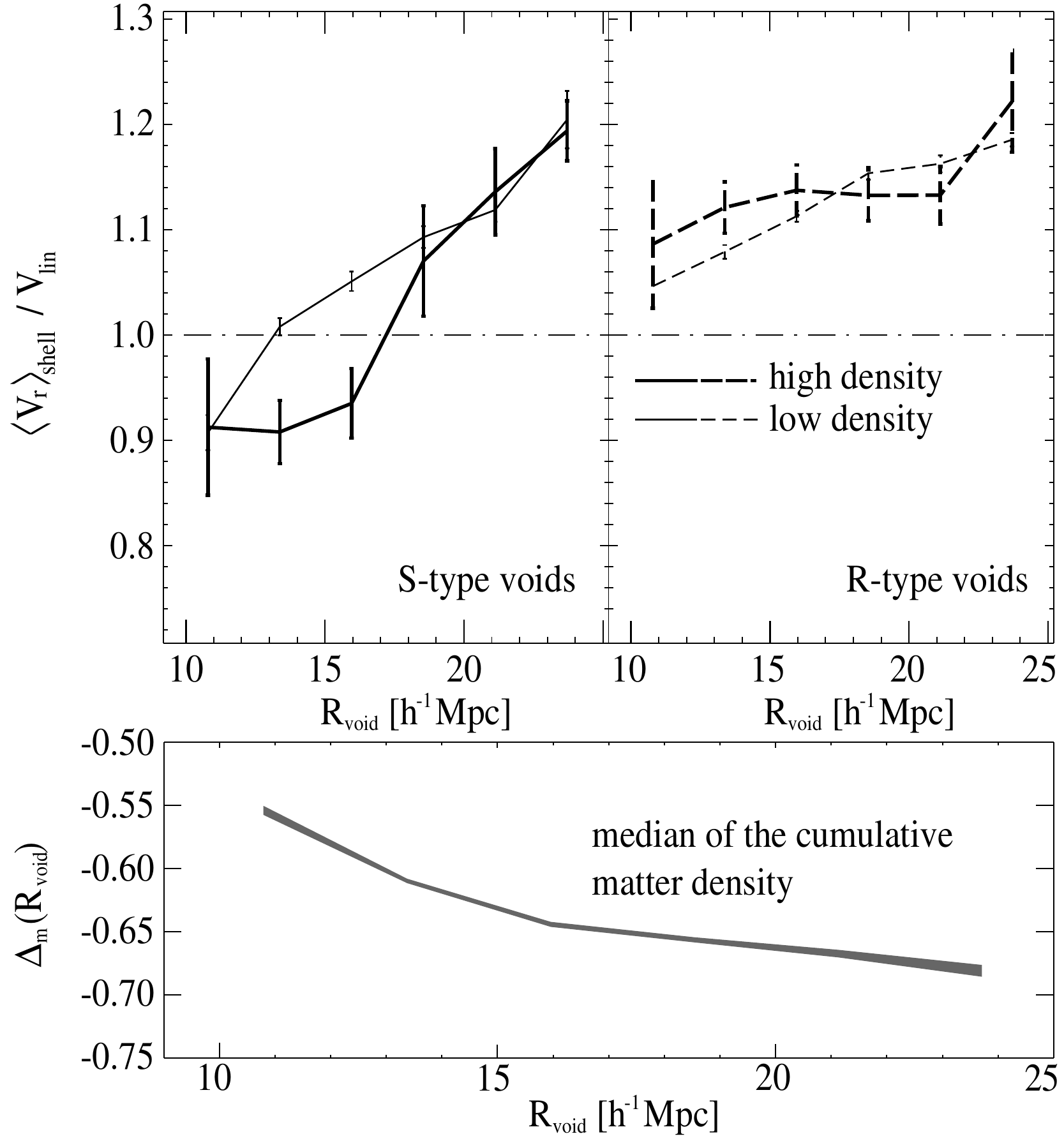}
\caption{
Median of the averaged radial velocities of each void relative to the linear
prediction as a function of the void radius for S-type (upper-left, solid
lines) and R-type (upper-right, dashed lines) voids.  The thick lines
correspond to the high density pixels and the thin lines to the low density
regions. Errorbars show the standard deviations. The lower panel shows the
median of the cumulative mass overdensity.  The dot-dashed horizontal line
indicates the prediction for radial velocities according to the linear theory.
All panels refers to the shells located at $r=R_{\rm void}$.
}
\label{fig:vr_rvoid}
\end{figure}

Finally, in Figure \ref{fig:vr_rvoid} we show the median of $\left< V_{\rm
r}\right>_{\rm shell}$ normalized to the linear prediction $V_{\rm lin}$ as a
function of the void size $R_{\rm void}$ for S-type voids (top-left panel,
solid lines) and R-type voids (top-right panel, dashed lines). These values
correspond to the shell located at a distance of one void radius from the
center.
The values for high density pixels are plotted in thick lines and the low
density regions are shown whit thin lines. 
It is clear from this Figure that there is a dependence of the mean radial
velocity of the shell on the void size, in both types and densities. This
dependence is more pronounced for S-type voids, and the high density regions
have more dispersion than the low ones.
The normalization with the linear model ensures that this correlation is a
feature beyond the linear dependence with size present in Eq. (\ref{eq:vlin}). 
The values of the median radial velocity for S-type voids are underpredicted by
a $\sim 10$ per cent in the median with respect to the linear theory for small
voids (and more evident for the high density pixels) and overpredicted by a
$\sim 20$ per cent in the median for the larger voids in our sample. 
For the sample of R-type voids we note an overprediction from $\sim 10$ to
$\sim 20$ per cent in the median in all the sizes range.
At the bottom panel of Figure \ref{fig:vr_rvoid} we show the median of the
cumulative mass overdensity inside the shell located at $R_{\rm void}$ as a
function of the void radius for the whole sample, irrespective of the void
environment.
As can be seen, there is a trend with the void radius which can be used to
explain the increasing deviations from linear theory with void size. That is,
larger voids tend to have more non-linear values of cumulative mass
overdensity, which as a result make the dynamic of these voids less described
by the linear theory. 
However, the differences between the deviations when comparing R- and S-type
voids are notably different.
This result becomes more relevant if we notice that the behaviour of the
cumulative overdensity with void size have negligible difference for both R-
and S-type voids (the corresponding curves can not be seen due to the fact that
they lie inside the grey shaded area in the bottom panel of Figure
\ref{fig:vr_rvoid}).
Therefore, even when the discrepancies with linear theory are mostly explained
by the emptiness of the inner parts of underdense regions, they exhibit a
notably different behaviour depending on the large scale environment.

\subsection{Structures in the high density regions}
\label{sec:structures}

Once pixels have been classified according to its particle density contained
within the spherical shell, we can separate them in density intervals in order
to study their properties.
The selected pixels in a density layer form clustered regions or
``chunks'', which represent the projection of structures surrounding the void
within a spherical slice of a given size.
These structures appear in a rich variety of shapes and complexity, which
depend on the density of their constituent pixels, the distance to the void
centre and the dynamics of the corresponding void.
We argue that these dependences encode useful information about the
large--scale structure of the void--filament network, and also possibly about
their formation and dynamics.
We aim to study the morphological properties of the pixel structures in order
to seek for and quantify such dependencies.
As a first step, we isolated the structures by forming groups of adjacent or
near pixels in the same density range.
This can be achieved by applying a FoF percolation algorithm to the set of
pixels, with an angular linking length equal to the mean pixel separation for
the complete void sample divided by the void radius.
The mean separation is around to $\sim 3\hmpc$ for the case of the densest
pixels.
This procedure leads to a number of pixel groups with a diversity of
arrangements, from compact to multibranch filamentary or "spider like"
structures.
A number of strategies could be used in order to quantify the geometry of the
arrangements of pixels.
The simplest method to characterize their forms is possibly the computation of
a normalized inertia tensor to obtain the elongation as the ratio of the
minor--to--major axes.
The semi--axis method has been used to characterize both 3D and 2D structures
\citep[][and references therein]{babul_quantitative_1992,
luo_three_1995,sathyaprakash_morphology_1998,Paz_2006}.
This procedure, however, does not distinguish between different levels of
compactness or filamentarity.
Other approaches have been proposed to characterize the geometry of complex
structures, most notably
Minkowsky functionals \citep{bharadwaj_evidence_2000,
basilakos_shape_2003,einasto_richest_2007, costa-duarte_morphological_2011},
surface modeling via triangulated networks \citep{sheth_measuring_2003} and
shapefinders of several kinds \citep[e.g.
][]{sahni_shapefinders_1998,aragon_multiscale_2007}.

We find useful to represent the set of pixels in a group by a two-dimensional
graph, where each node is a pixel and any pair of nodes is connected by an edge
with a weight equal to the angular distance between the centers of the pixels
corresponding to the adjacent nodes.
The disposition of nodes and the general geometry can be easily computed by
means of the minimal spanning tree of the graph \citep[][hereafter
MST]{barrow_msp_1985}, defined as the shortest weighted path that connects all
the nodes without cycles.
This tree can be constructed for a set of pixels with the Kruskal algorithm
\citep{kruskal_mst_1956}, using the positions of pixels and the spherical
distances between them.
An important MST characteristic is its length $L$, equal to the sum of the weights
of all its edges.
The maximum distance between any pair of nodes, $D_{\rm max}$, is also a useful
quantity, which combined to the length allows to define an estimator of
structure elongation, $E=D_{\rm max}/L$.
This last parameter is dimensionless and by construction its value ranges
between 0 and 1, where groups with an elongation value of roughly 1 are
filaments or thread--like structures, and groups with a low elongation value
are compact or concentrated structures.

We compute the maximum separation, length and elongation estimators for all
structures arising from pixels in the high density layer, once again as in
previous subsections, in a shell centered at $r=R_{\rm void}$. 
As we discussed in section \ref{sec:shells}, we are interested in characterizing
in a statistically robust basis, the deviations from linear theory in the
angular structure around voids, and whether they are related or not with the
void large scale environment (i.e. R- or S-type void).

\begin{figure}
\centering
\includegraphics[width=0.45\textwidth]{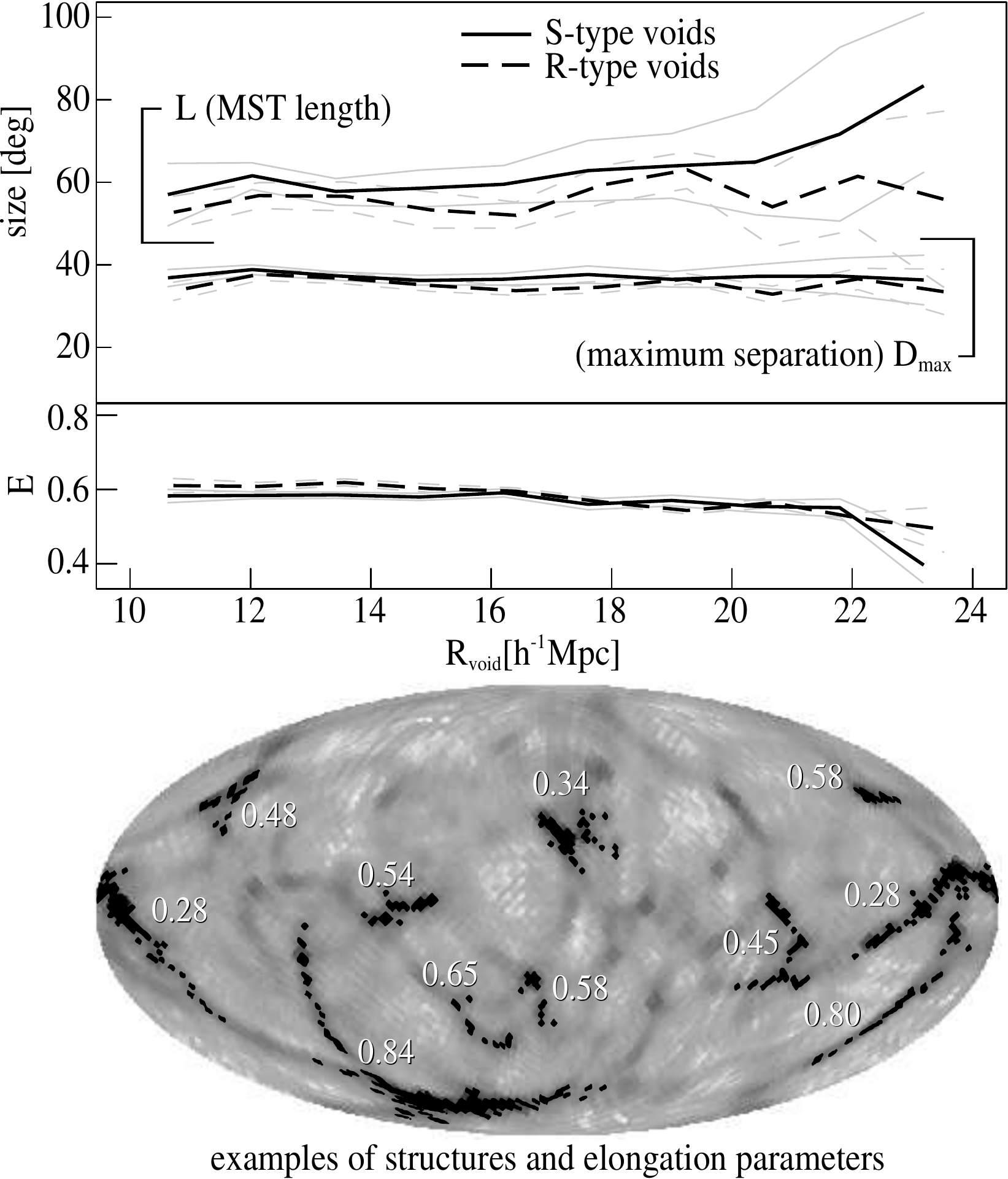}
\caption{
Shape parameters of structures within the shells centered at $R_{\rm void}$, of
S-type (solid lines) and R-type (dashed lines) voids.  Median values are shown
in black and the corresponding errors in grey. In the upper panel we show the
minimal spanning tree length ($L$) and the maximum separation ($D_{\rm max}$).
In the middle panel, the median of the elongation values $E=D_{\rm max}/L$ are
given as a function of void radius. In the lower panel we show an example of
the angular distribution of structures and their corresponding elongation
parameter values.
}
\label{fig:shape_rvoid}
\end{figure}

In Figure \ref{fig:shape_rvoid} we show the median values (black lines) and the
standard deviation (grey lines) of MST parameters as a function of the void
radius. S-type voids are drawn in solid lines and R-type voids in dashed lines.
In the top panel we show the angular size of the MST length ($L$) and the
maximum separation ($D_{\rm max}$), and in the middle panel the elongation
parameter $E=D_{\rm max}/L$. 
As it can be seen in this Figure, these parameters do not depend significantly
on void size, indicating that the definition of these quantities is not biased
by the spatial size of the void shell.
In order have a visual insight on the structure shapes and the elongation
values, we show in the lower panel an example of the distribution of structures
and their elongation parameters in a void shell.

In the next Section we use these parameters to explore the dynamics of
structures depending on the void classification and size.


\subsection{Dynamics of structures}
\label{sec:structures_dynamics}

Since the mass in the shells around voids is arranged in
structures, specially when density levels are considered, the study of the
velocities of these structures merits further investigation.
In Section \ref{sec:structures} we described the procedure to characterize the
shapes of groups of pixels, as a proxy to the projection of structures within
the shell, by means of the minimal spanning tree.
This approach has a number of advantages, mainly its simplicity and the ability
to differentiate between filamentary structures with several degrees of
complexity.

In this Section we aim to characterize the smoothness of the radial flow of
structures. 
To do this, we compute the averaged void--centric radial velocities,
$\left<V_{\rm r} \right>_{\rm str}$, and the averaged tangential velocities,
$\left<V_{\rm t}\right>_{\rm str}$, of all the structures resulting from the
FoF percolation of high density pixels on each shell.
The smoothness of the radial flow can be quantified by measuring the median of
the ratio of radial to tangential velocities, namely $\left<V_{\rm r}
\right>_{\rm str} / \left<V_{\rm t}\right>_{\rm str}$.

\begin{figure}
\centering
\includegraphics[width=0.45\textwidth]{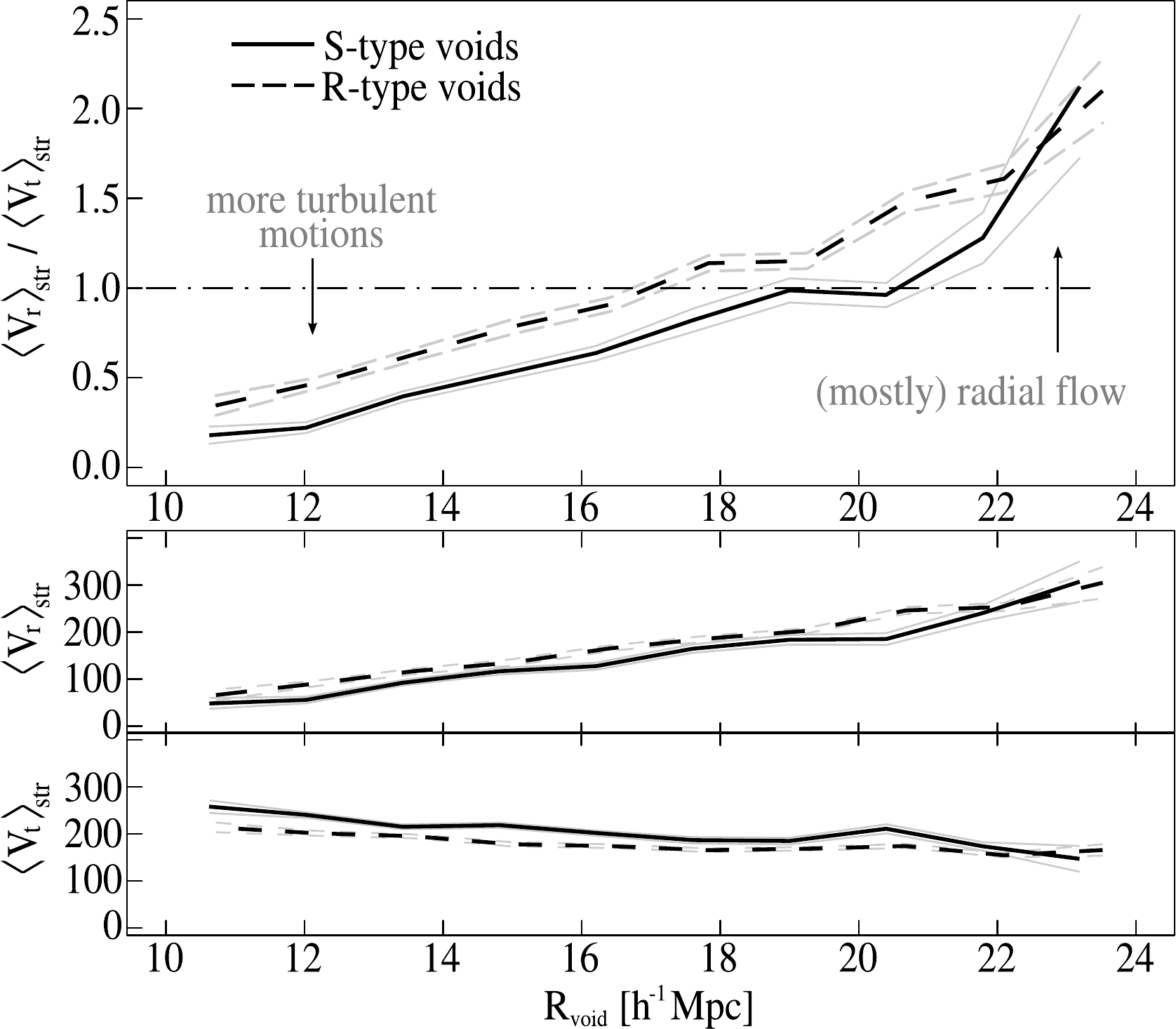}
\caption{
Median of the mean radial (middle panel) and tangential (bottom panel)
velocities, in $\kms$, of structures of S-type (solid lines) and R-type
(dashes lines) voids, as function of void radius.  On the upper panel we show
the ratio of radial-to-tangential velocity, which gives an estimate of the
mass flow \textit{smoothness}. Grey lines indicate the errors of the medians.
The dot-dashed line mark the transition from more turbulent motions to a radial
flow.
}
\label{fig:smoothness}
\end{figure}

In Figure \ref{fig:smoothness} we show the results of $\left<V_{\rm r}
\right>_{\rm str}$, $\left<V_{\rm t}\right>_{\rm str}$ and $\left<V_{\rm r}
\right>_{\rm str} / \left<V_{\rm t}\right>_{\rm str}$ as function of void
radius considering separately S- and R-type voids.
It can be seen that the radial flow smoothness of structures in void shells
increases with void radius so that the largest the voids, the more significant
the outward void--centric radial velocity as compared to their tangential
motions. We find structures in R-type voids to have systematically smoother
radial outflows compared to structures in S-type voids, and that radial flows
dominate the dynamics for voids larger than $\sim 18\hmpc$. Besides, it can be
seen in the lower panels that this behavior is caused by the combined effect of
structures in S-types to have a systematically lower radial flow and larger
tangential motions.
These results could be useful when modeling the velocity field in voids.  For
instance, in the case of measurements in redshift space (either correlation or
stacking statistics), for certain void sizes, the results would be more
affected by uncertainties in the pair-wise velocity dispersion than in the mean
flow velocity. This could be inferred from the mentioned above dominance of
tangential velocities on voids with radius below $18\hmpc$. This is also
the case for S-type voids at all sizes: they seems to be more affected by
tangential velocities than R-type regions.

\begin{figure}
\centering
\includegraphics[width=0.45\textwidth]{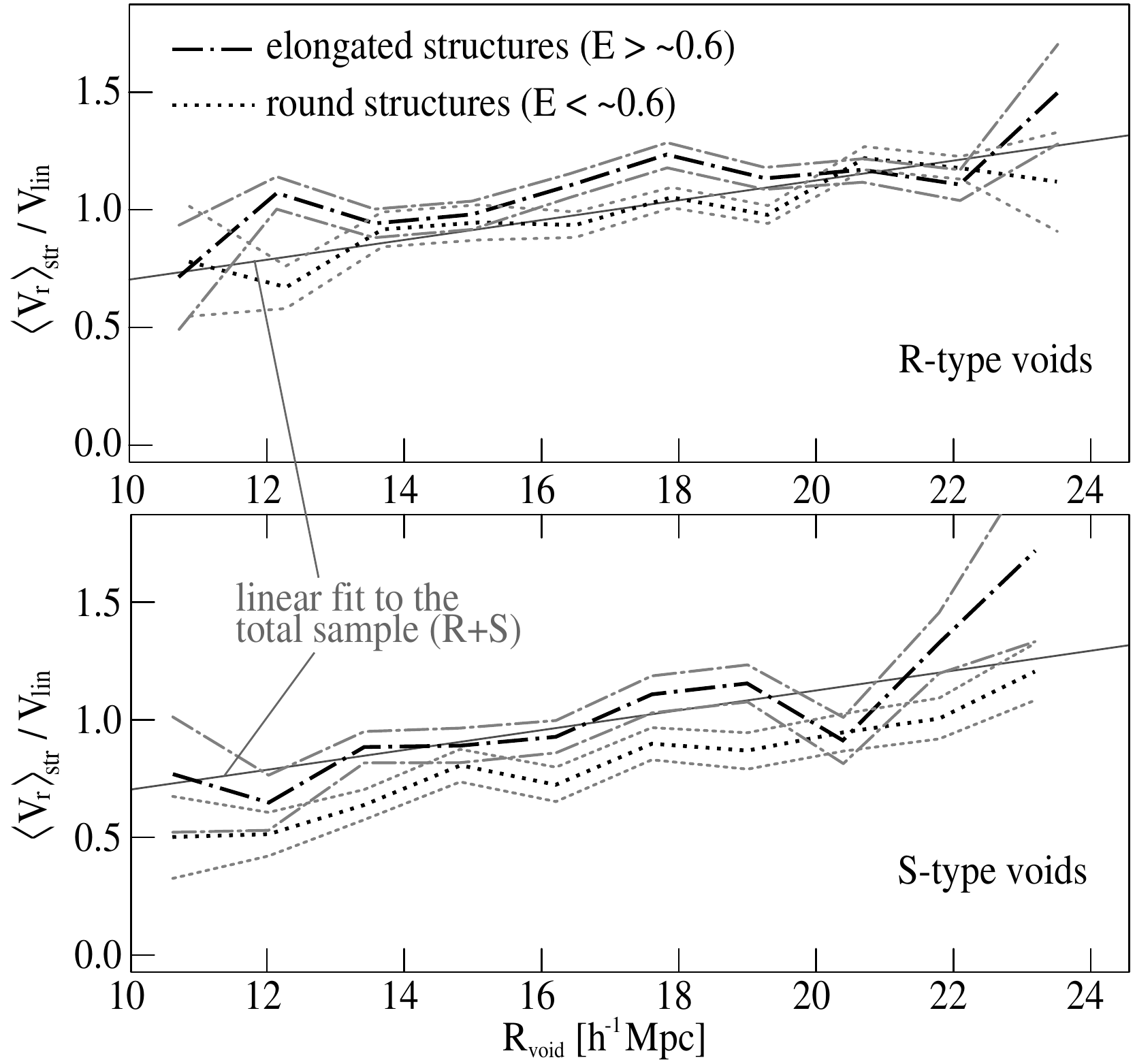}
\caption{
Median of the mean radial velocities of structures of each void relative to the
linear prediction of Eq. (\ref{eq:vlin}) for R-type (upper panel) and S-type
voids (lower panel) as function of void size.  The samples has been divided
into elongated (dot-dashed lines) and round structures (dotted lines), as
described in the text. The trend for the whole sample (without separating in
elongation or void type) is showed with the thin solid line. Grey lines
correspond to the standard deviation.
}
\label{fig:vrvl_rvoid}
\end{figure}

We have additionally explored the behavior of the mean radial velocities of
structures relative to the linear prediction of Eq. (\ref{eq:vlin}) as function
of $R_{\rm void}$. The results are given in Figure \ref{fig:vrvl_rvoid} 
where we have considered separately the sample of structures identified in R-type
(upper panel) and S-type voids (lower panel), and each of this subsamples into
elongated (dot-dashed lines) and round (dotted lines) structures. 
This last separation was performed according to the results of the middle panel
of Figure \ref{fig:shape_rvoid}: structures with elongation values higher than
the mean ($E > \sim 0.6$) are considered as elongated, and structures with
lower values as round. 
The solid thin line represents the global trend of the whole sample (a linear
fit to the total void sample), with no shape restrictions.
It can be seen that for both R- and S-type voids, the linear prediction
provides an overestimation of the radial velocity of about $\sim 20$ per cent
in the median for small voids and an underestimation of $\sim 20$ per cent in
the median for the larger voids.
The global behaviour for S-type voids in Figure \ref{fig:shape_rvoid} agree
with the trend shown in the left panel of Figure \ref{fig:vr_rvoid}. 
However, for R-type voids we note a discrepancy since the radial motion of
shells is always underestimated by the linear model (see right panel of Figure
\ref{fig:vr_rvoid}), while structures identified in the high density pixels
present the same velocity trend than structures in S-type voids. 
Depict this qualitative similar behaviours in both Figures
(\ref{fig:vrvl_rvoid} and \ref{fig:vr_rvoid}), it should be emphasized
that the quantities shown on each case have a different meaning.  For
instance, results on Figure \ref{fig:vr_rvoid} can be thought as the
typical error obtained when the mean velocity of a given shell is
predicted by linear theory. Meanwhile, in Figure \ref{fig:vrvl_rvoid},
the displayed trend indicate the typical difference observed between
the actual velocity of a given structure with the mean predicted by the
linear model.  

%

\section{Conclusions}
\label{sec:conclusions}

Based on the dynamical dichotomy first predicted by
\citet{sheth_hierarchy_2004}, and confirmed on observations in previous
works (Paper I, Paper II), we have explored the interplay between the inner
dynamics of voids and its large scale environment. 
The presented study goes beyond our previous analysis, that examined the mean
velocity profile of voids, here we have studied the velocity field structure
(higher moments, angular distribution, etc.) in shells around them. 
The aim of this analysis was not only to inquire the existence of correlations
between the inner void dynamics and its environment, but also was to look for
clues on the origin of the nonlinearities \citep[Paper
II,][]{hamaus_profile_2014} in the velocity field of voids.

We have analysed voids in different dynamical regimes, according to the R/S
type classification presented in Paper I.
In order to accomplish this, we have considered spherical shells around voids at
several void--centric distances, analysing the transition between the domains
of voids and surrounding structures.
An statistical study on the spatial and velocity distributions on void-centric
shells shows notably differences between void types and sizes.
At the inner parts of the utmost S-type voids, the velocity distribution takes
the form of angular patches with predominant infall or outflow radial
velocities, even though the mean velocity in all cases is outwards.
Meanwhile, for R-type voids, the velocity angular distribution is consistent
with a pure outflow pattern. 
As a result, the distributions of deviations from the bulk flow on R-type voids
are typically narrow, well centered on its mean values. 
In contrast, S-type voids typical exhibits an inner broad distribution of
radial velocities, without a neat bulk flow peak or, in some cases, incipient
bimodality.
It should be noticed that between both utmost behaviours there is a continuous
range of velocity patterns.

In the analysis for the radial velocity field at shells in increasing radii we
find that R-type voids the velocity field remains with positive outflow at
different shell radii, which contrast with the behaviour of S-type voids that
show a significant infall beyond $1.5$ void radii, in agreement with Paper II.
The spread of void--centric radial velocities for S-type voids is significantly
larger than that of R-type voids, particularly at large distances from the center. 
We also found that the distribution of velocities of the high density regions
exhibit an infalling tail, which is remarkable for S-type voids but also
notably on R-type voids.
The linear model provides a suitable fit to the observed radial infall/outflow;
however, in S-type voids, there is a systematic offset towards smaller values
by approximately $\sim 30-50\kms$.
We have also seen that linear theory successfully predicts Gaussian deviations
on radial velocity fluctuations (i.e. skewness and kurtosis moments are
negligible on linear theory deviations).
The results summarized in the above paragraphs, could be important when
modeling the pair-wise velocity function on redshift space measurements and
models (see for instance Paper II). 

The ability of the linear theory is also tested when comparing the ratio of the
observed radial velocity to the predicted one in the shell located the void
radius, founding that this ratio is an increasing function of the void size for
both S- and R-type and high and low density regions.
An explication of this trend can be found by analysing the cumulative mass
overdensity inside the shell, $\Delta_{\rm m}(R_{\rm void})$. 
As expected, larger voids have stronger non linear values of mass density,
therefore the dynamics of larger voids are less described by a linear
approximation \citep[in agreement for instance with][and references
therein]{hamaus_profile_2014}.
However, it is important to note that although the values of $\Delta_{\rm m}$
are indistinguishable between both S- and R-type voids, the trend of radial
velocity with void size observed presents notably differences depending of the
large scale environment.
These results reinforce the idea repeated along this work: the nonlinearties
observed in the inner parts of voids could arise from a coupling of scales.
Large scale environment defined arround $3R_{\rm void}$ has a significant
correlation with dynamics at inner scales ($-0.8<\Delta_{\rm m}<-0.6$).

The structures defined in the high density regions of void shells provide
further dynamical studies. 
We find that these structures shown similar deviations from linear theory than
previous results, that is larger departures are seen at larger sizes.
We found again that velocities traced by structures in both S- and R-type have
a common transition scale, from linear model overprediction to underprediction,
at void sizes around $\sim 18\hmpc$.
Besides, elongated structures exhibit systematically larger $\left< V_{\rm
r}\right>_{\rm str}/V_{\rm lin}$ values with respect to the rounder structures
by $\sim 20$ per cent in the median.
We have analysed the radial to tangential velocity ratio of the structures as a
suitable measure of the void-centric flow smoothness. 
The results of this study show an increasing trend of radial flow smoothness as
a function of void size produced by a systematically increasing mean radial
velocity and a continuously decreasing tangential velocity. 
The observed correlation suggests that large voids (with sizes greater than
$R_{\rm void} \sim 18\hmpc$) have more frequently boundaries with structures
dominated by radial outflows, in contrast with a more turbulent motion of the
structures at boundaries of small voids.
We also find structures in R-type voids to have systematically smoother radial
outflows by $\sim 25$ per cent in the median compared to structures in S-type
voids.

As a final remark, is worth to notice that R-type voids with larger sizes are
the regions best described by linear theory and dominated by radial outflows.
In addition, low density structures seem to trace the inner velocity field in
better agreement with linear theory. 
This could be important when designing cosmological tests based on voids
samples and the oncoming next generation of large volume galaxy surveys as
HETDEX \citep{hill_hetdex_2008}, Euclid \citep{euclid_2011}, SDSS-III
\citep{sdss3_2011} and VIPER \citep{2014_viper_Micheletti}.
In particular measurements which depends on a proper modeling of the velocity
or density void profiles \citep[e.g. the Alcock-Paczynski test or the
Integrated Sachs-Wolfe effect,][]{ 2008ApJ...683L..99G,2011ApJ...732...27P,
2012Lavaux,sutter_first_2012, hernandez-monteagudo_signature_2013,
2013A&A...556A..51I, 2014ApJ...786..110C, 2014MNRAS.439.2978C}. 
For these next generation surveys, given the large volume sampled, it is
expected to identify voids by using high luminosity tracers. 
Voids catalogues obtained from such galaxy samples are expected to have
predominantly R-type underdense regions.  
Even though, these type of regions behaves on better agreement with linear
theory, the brighter sample of galaxies is expected to trace the velocity field
of high density structure, which has shown in this work have systematic departures
from linearity at the inner regions. 
Therefore, it could be required a proper model of the velocity expansion profile of
this systems when measurements are performed in redshift space.
In the same direction, studies based on low redshift samples, as the main
galaxy sample of the SDSS, can be benefited by an appropriate treatment of the
velocity tracer population for both S- and R-type underdense regions.


\section*{Acknowledgments}
We thanks helpful comments and suggestions from the referee Paul Sutter, which
have improved substantially the work presented. This work was partially
supported by the Consejo Nacional de Investigaciones Cient\'{\i}ficas y
T\'ecnicas (CONICET), and the Secretar\'{\i}a de Ciencia y Tecnolog\'{\i}a
(SeCyT), Universidad Nacional de C\'ordoba, Argentina. 
ANR acknowledges receipt of fellowships from CONICET. 
Plots are made using {\sc R} software and post-processed with {\sc Inkscape}.

\bibliographystyle{mn2e}
\bibliography{references}

\end{document}